\documentclass[
reprint, 
 amsmath,amssymb,
 aps, prl,twocolumn,
]{revtex4-2}

\usepackage{subfigure}
\usepackage{caption2}
\usepackage{graphicx}
\usepackage{dcolumn}
\usepackage{bm}
\usepackage{hyperref}
\usepackage[mathlines]{lineno}

\usepackage{slashed}
\begin{document}


\title{\textbf{Possible bound states in the triple-$\eta_c$ and triple-$J/\psi$ systems} 
}%

\author{Guo-Feng Xu$^1$}
\author{Xu-Liang Chen$^1$}
\author{Jin-Peng Zhang$^1$}
\author{Ning Li$^1$}
\email{lining59@mail.sysu.edu.cn}
\author{Wei Chen$^{1,\, 2}$}
\email{chenwei29@mail.sysu.edu.cn}

\affiliation{$^1$School of Physics, Sun Yat-sen University, Guangzhou 510275, China \\ 
$^2$Southern Center for Nuclear-Science Theory (SCNT), Institute of Modern Physics, 
Chinese Academy of Sciences, Huizhou 516000, Guangdong Province, China}

\begin{abstract}
The observations of fully-charm tetraquark states in the LHCb, CMS and ATLAS experiments suggested the existence of the hadronic molecules of two-charmonium states, which may also imply  bound states in the three-charmonium systems. In this work, we study the possible bound states in the triple-$\eta_c$ and triple-$J/\psi$ systems with $J^{PC}=0^{-+}$ and $1^{--}$, respectively. In QCD sum rules, we calculate the two-point correlation functions and spectral functions up to the dimension-four gluon condensate. We use the iterative dispersion relation approach to deal with the five-loop banana integrals, which significantly improves the computational efficiency. Our results show that the masses of triple-$\eta_c$ and triple-$J/\psi$ states lie below the corresponding mass thresholds, supporting the existence of such three-body bound states. 
\end{abstract}

\maketitle


\textit{\textbf{Introduction.}} \textemdash \,
Over the past half-century, the search for multiquark states has always been a very intriguing research topic in hadron physics.  Since 2003, there has been a resurgence of such investigations following the observation of numerous good candidates, such as the hidden-charm pentaquarks $P_c/P_{cs}$,the doubly-charm tetraquark $T_{cc}^+$, the fully-charm tetraquark $X(6900)$, and so on~\cite{ParticleDataGroup:2024cfk,Chen:2016qju,Guo:2017jvc,Liu:2019zoy,Brambilla:2019esw,Guo:2019twa,Chen:2022asf,Liu:2024uxn,Zhu:2024swp,Wang:2025sic}.


In 2020, the LHCb Collaboration reported the $X(6900)$ state and a broad structure between $6.2-6.8$GeV in the $J/\psi J/\psi$ invariant mass spectrum~\cite{LHCb:2020bwg}. Three years later, the ATLAS and CMS Collaborations confirmed the existence of $X(6900)$ in the same channel~\cite{ATLAS:2023bft,CMS:2023owd}. In addition, ATLAS also reported structures $X(6400)$ and $X(6600)$~\cite{ATLAS:2023bft}, while CMS reported $X(6600)$ and $X(7200)$ in the $J/\psi J/\psi$ invariant mass spectrum~\cite{CMS:2023owd}. Meanwhile, the existence of $X(6900)$ and $X(7200)$ was also confirmed by ATLAS in the $J/\psi\psi(2S)$ final states~\cite{ATLAS:2023bft}. Observed in the $J/\psi J/\psi$ and $J/\psi\psi(2S)$ channels, these resonance structures have been extensively considered as good candidates for the fully-charm tetraquark states~\cite{Chen:2016jxd,Bedolla:2019zwg,Albuquerque:2020hio,Giron:2020wpx,Guo:2020pvt,Karliner:2020dta,Liu:2019zuc,Wang:2018poa,Wang:2019rdo,Yang:2020rih,Yang:2020wkh,Zhang:2020xtb,Zhu:2020xni,Cao:2020gul,Albuquerque:2021erv,Wang:2021mma,Wang:2021taf,Yang:2021zrc,An:2022qpt,Wu:2022qwd,Kuang:2022vdy,Lu:2023ccs,Feng:2023agq,Yang:2021hrb,Galkin:2023wox,Wu:2024euj,Chen:2024bpz,Chen:2022sbf,Biloshytskyi:2022pdl,Sang:2023ncm,Anwar:2023fbp,Agaev:2023rpj,Wang:2020gmd,Feng:2020qee,Maciula:2020wri,Ma:2020kwb,Wang:2020tpt,Gong:2020bmg,Feng:2023ghc,Yan:2023lvm,Yu:2022lak,Ke:2021iyh,Yu:2022lak,Niu:2022cug,Zhou:2022xpd,Kuang:2023vac,Liu:2021rtn,Lu:2023aal}, including the hadronic molecules composed of two charmonia~\cite{Agaev:2023rpj,Lu:2023ccs,Lu:2023aal}. In Refs.~\cite{Dong:2020nwy,Nefediev:2021pww,Dong:2021lkh,Agaev:2023ruu,Song:2024ykq}, the $X(6200)$ structure was predicted to exist as a bound/virtual state near the $J/\psi J/\psi$ threshold. 

If two charmonia can form a bound state, one may expect the existence of a three-charmonium state such as those in the triple-$\eta_c$ and triple-$J/\psi$ systems. In Ref.~\cite{Pan:2024ple}, the possible bound states and the Efimov effect~\cite{Efimov:1970zz,Naidon:2016dpf} of the triple-$J/\psi$ system were investigated by employing the Gaussian expansion method. They found a shallow bound triple-$J/\psi$ state even in the case that the attractive interaction between two $J/\psi$ mesons is weak. Recently, the CMS Collaboration observed the simultaneous production of three $J/\psi$ mesons in proton-proton collisions and measured the inclusive fiducial cross section~\cite{CMS:2021qsn}, which shed light on the production of such a three-body bound state. Moreover, there have been numerous studies on fully heavy hexaquark systems in both the hadronic molecule and compact state configurations~\cite{LQCD1,LQCD2,LQCD3,LQCD4,QM1,QM2,QM3,QM4,QM5,QM6,QM7,QM8,OBE,WZG}. However, the results obtained in these theoretical approaches are quite different from each other. More investigations are needed to deepen our understanding of the fully heavy hexaquark systems. In this work, we study the possible bound states of the triple-$\eta_c$ and triple-$J/\psi$ systems using the QCD sum rules method.

\textit{\textbf{Formalism.}} \textemdash \,
In this section, we introduce the QCD sum rules method for investigating fully heavy hexaquark states~\cite{sr1,sr2}. We construct the following two interpolating  currents
	\begin{eqnarray}
		\begin{split}
			J (x) =& (\bar{c}_a \text{i} \gamma_5 c_a) (\bar{c}_b \text{i}  \gamma_5 c_b) (\bar{c}_c
			\text{i}  \gamma_5 c_c) ,
			\\
			J_{\mu} (x) =& (\bar{c}_a \gamma_{\mu} c_a) (\bar{c}_b \gamma_{\nu} c_b) (\bar{c}_c
			\gamma^{\nu} c_c),
		\end{split}
	\end{eqnarray}
to study the triple-$\eta_c$ and triple-$J / \psi $ systems, respectively. The scalar current $ J(x) $ carries quantum numbers $J^{PC} = 0^{-+}$, while the vector current $J_{\mu}(x)$ carries $ J^{PC} = 0^{+-} , 1^{--}$. The two-point correlation functions induced by these two interpolating currents are defined as
		\begin{eqnarray}\label{correlators}
				\begin{split}
					 \Pi (q^2) =& \text{i}  \int d^4 x e^{\text{i}  q \cdot x} \langle 0 | T [J (x)
				J^{\dagger} (0)] | 0 \rangle ,
				\\
				\Pi_{\mu \nu} (q^2) = &\text{i} \int d^4 x e^{\text{i}  q \cdot x} \langle 0 |
				T [J_{\mu} (x) J_{\nu}^{\dagger} (0)] | 0 \rangle
				\\
				= & \left( \frac{q_{\mu} q_{\nu}}{q^2} - g_{\mu \nu} \right) \Pi_1 (q^2)
				+ \frac{q_{\mu} q_{\nu}}{q^2} \Pi_0 (q^2),
			\end{split}
	\end{eqnarray}
    in which the invariant functions $\Pi_1(q^2)$ and $\Pi_0(q^2)$ correspond to the $1^{--}$ and $0^{+-}$ pieces, respectively. We shall extract the invariant function $ \Pi_1(q^2) $ to investigate the vector triple-$J / \psi$ state.  
In this work, we don't study the $0^{+-}$ state due to the external $P$-wave excitation in $J_\mu(x)$, which may lead to a higher state in this channel. 

In general, the interpolating currents can couple to all structures with the same quantum numbers, including bound states, tri-meson scattering states and continuum. All of them will contribute to the correlation functions in Eq.~\eqref{correlators}. However, it has been demonstrated that only exotic multi-quark states can saturate the QCD sum rules, but not the scattering states. Moreover, the contributions from the scattering states are numerically small~\cite{Wang:2025sic,Wang:2020cme,Wang:2019gal,Albuquerque:2021tqd,Wang:2020fuh}. 

    \begin{figure}[htbp]
    \centering
    \subfigure[]{
        \includegraphics[width=0.14\textwidth]{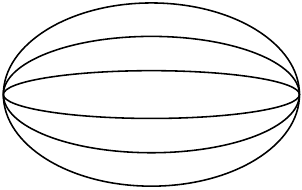}
        \label{fig:pert}
    }
    \subfigure[]{
        \includegraphics[width=0.14\textwidth]{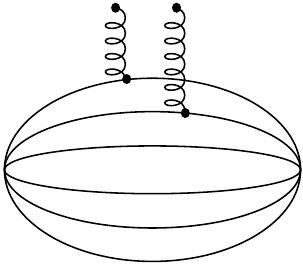}
        \label{fig:GG1}
    }
    \subfigure[]{
        \includegraphics[width=0.14\textwidth]{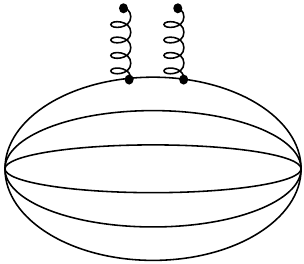}
        \label{fig:GG2}
    }
    \caption{Feynman diagrams involved in the OPE series. }
    \label{fig1}
\end{figure}

At the quark-gluonic level, the correlation functions can be calculated via the method of operator product expansion (OPE). We adopt the following heavy quark propagator
	\begin{eqnarray}
    \begin{split}
		 \text{i} S_Q^{i j} (p) &= \frac{\text{i} \delta^{i j}}{\slashed{p} - m_Q} 
          + \frac{\text{i} \delta^{i j}}{12} \langle
		g_s^2 G G \rangle m_Q \frac{p^2 + m_Q \slashed{p}}{(p^2 - m_Q^2)^4} 
        \\
          &-
		\frac{\text{i}}{4} g_s \frac{\lambda^a_{ij}}{2} G^a_{\mu \nu} \frac{\sigma^{\mu
				\nu} \left( \slashed{p} + m_Q \right) + \left( \slashed{p} + m_Q \right)
			\sigma^{\mu \nu}}{(p^2 - m_Q^2)^2} .
        \end{split}
	\end{eqnarray}
	where $\lambda^a_{ij}$ is the Gell-Mann matrix, with $i, j$ the color indices. In the fully-heavy systems, only gluon condensates contribute to the nonperturbative effects, and thus we consider the OPE series up to dimension four condensate
    \begin{eqnarray} \label{OPEseries}
		 \Pi^{\text{OPE}} (q^2) = \Pi_a^{\text{Pert}} (q^2) + \Pi_b^{\langle G^2\rangle} (q^2) +  \Pi_c^{\langle G^2\rangle} (q^2),
	\end{eqnarray}
which are depicted in Fig.~\ref{fig1}  (a-c), respectively. In this work, we shall not consider the dimension six tri-gluon condensates in the OPE series, which are usually negligible in fully heavy multiquark systems~\cite{Chen:2016jxd,Wang:2020avt,Wang:2021mma,Wang:2021taf,Yang:2021zrc,Wu:2021tzo}. Moreover, the computations of tri-gluon condensates will involve the massive high-power propagators in the 5-loop integrals, which are very difficult and complicated at the present stage. Actually, the calculations are already challenging enough for the Feynman diagrams in Fig.~\ref{fig1} (a-c). We shall introduce some integral techniques in the following to improve the computational efficiency. 

At the hadronic level, the hadronic spectral function can be parametrized by the ``narrow resonance'' assumption
	\begin{eqnarray}
	\rho^{\text{PH}} (s) \equiv \pi^{- 1} \text{Im} \Pi^{\text{PH}} (s) =f_X^2 \delta (s - m_X^2) + \cdots,
	\end{eqnarray}
	where $m_X$ and $f_X$ are respectively the hadron mass and coupling of the ground state, and \textquotedblleft$\cdots$\textquotedblright \, represents contributions from the continuum/excited states. The coupling constant $f_X$ is defined as 
\begin{eqnarray}
\begin{split}
        \left\langle 0 \left\lvert J\right\rvert X\right\rangle &=f_X\,,
        \\
        \left\langle 0 \left\lvert J_{\mu}\right\rvert X\right\rangle &=f_X\epsilon_{\mu}\,,
\end{split}
\end{eqnarray}
in which $\epsilon_{\mu}$ is the polarization vector.
    
Applying the dispersion relation (DR), one obtains the correlation function at the hadronic level
	\begin{eqnarray}
		 \Pi^{\text{PH}} (q^2) =  \int_{s_N}^{\infty} d s \frac{ (q^2)^n\rho^{\text{PH}} (s)}{s^n (s - q^2 -\text{i}0^+)} + \sum^{n - 1}_{i = 0} b_i
		(q^2)^i ,
	\end{eqnarray}
where $s_N = (6m_Q)^2$ is the physical threshold and $b_i
(q^2)$ is the unknown subtraction term. Based on quark-hadron duality, one can establish the sum rules after performing the Borel transform on  $\Pi^{\text{OPE}}(q^2)$ and $\Pi^{\text{PH}} (q^2)$ 
    \begin{equation}
\begin{aligned}
\mathcal{L}_k(s_0, M_B^2) &\equiv f_X^2 \, (m_X^2)^k \, e^{-m_X^2/M_B^2} \\
                          &= \int_{s_N}^{s_0} s^k \, \rho^{\text{OPE}}(s) \, e^{-s/M_B^2} \, ds,
\end{aligned}
\end{equation}
	where $s_0$ is the continuum threshold parameter and $M_B$ is the Borel mass. Thus, the mass of the hadronic ground state $m_X$ can be determined as
	\begin{eqnarray}
		m_X^2 = \frac{\int_{s_N}^{s_0 } d s \rho^{\text{OPE}} (s)s e^{- s/M_B^2 }}{{\int_{s_N}^{s_0 } d s \rho^{\text{OPE}} (s) e^{- s/M_B^2 }}}.
	\end{eqnarray}

As shown in Fig.~\ref{fig1}, the evaluations of the OPE series involve the massive five-loop banana integrals, whose calculations are extremely slow and resource-consuming in the Feynman parameterization method. In our calculations, we shall employ an iterative dispersion relation (IDR) method to improve the computational efficiency~\cite{Govaerts:1984hc,Remiddi:2016gno,Chen:2024ppj,Freitas:2016sty,Bauberger:2019heh,Freitas:2016zmy,Aleksejevs:2018tfr}. As an example, we illustrate the IDR method for a scalar integral with the general form
\begin{eqnarray}
		B_{\vec{n}} (q^2) \equiv \int  \left( \prod^5_{i = 1} \frac{d^D k_i}{(2
			\pi)^D} \frac{1}{D_i^{n_i}} \right) \frac{1}{D_0^{n_0}},
\label{scalarintegtal}
\end{eqnarray}
where $D_i  = k_i^2 - m_Q^2$, $D_0 = \left( q - \sum^5_{i = 1} {k_i} 
\right)^2 - m_Q^2$ are the inverse propagators, $n_i$ is the corresponding power, $\vec{n} \equiv
(n_0, n_1, \cdots,
n_5)$ is the power vector and $q$ is the external momentum. 
After completing the integral of $ k_1 $ in Eq.~\eqref{scalarintegtal}, we compute the imaginary part of the first loop and apply DR to rebuild the integral. The propagator-like term $ (q^2 - s)^{-1} $ in DR will then provide the new $D_0$ for the next-loop calculation. Terms without propagator-like structures do not contribute to the imaginary part in subsequent loops and thus can be omitted. All loop momenta $k_1$ to $k_5$ can be integrated in such a process. This approach transforms the multiloop computations into an iterative process of one-loop bubble integrations, thereby significantly enhancing the computational efficiency.
    
The DR can be safely applied  to the perturbative contribution in Fig.~\ref{fig:pert}, where all the involved propagator powers satisfy $n_i=1$. In Fig.~\ref{fig1}(b-c), the high-power propagators in $\Pi_{b/c}^{\langle G^2\rangle} (q^2)$ can break the applicability of DR. If $n_0 + n_1 \geq 4$ in a certain loop, the application of DR will lead to a divergent result. The calculations for integrals with tensor structures (tensor integrals) exhibit very similar properties. In such cases, the generalized dispersion relation (GDR) should be applied to deal with the small-circle subtraction problem and provide convergent calculations~\cite{Chen:2024ppj}. One can verify this easily by computing the one-loop bubble diagram. Given that the GDR representation for an integral is analytically challenging and computationally expensive, we employ the following techniques in practice to circumvent the need for GDR in our calculations of Fig.~\ref{fig1}(b-c). 

To calculate $\Pi_{b}^{\langle G^2\rangle} (q^2)$ in Fig.~\ref{fig:GG1}, one should confront the integrals with the propagator power vector $\vec{n} = (2, 2, 1, 1, 1, 1)$. If one integrates the inner momentum $k_1$ in the first loop, it is obvious that DR fails for such integrals, and  GDR must be mandatorily applied. By reordering the integration sequence, for example, integrating $k_2$ first (involving $ D_0 $ and $ D_2$) followed by $ k_1$ — the propagator power configuration transforms into $\vec{n} = (2, 1, 2, 1, 1, 1) $ satisfying $ n_0 + n_i < 4 $ at each loop. This modified configuration allows DR to be applicable, eliminating the need for GDR. 

Unfortunately, this reordering strategy is ineffective for $\Pi_{c}^{\langle G^2\rangle} (q^2)$ in Fig.~\ref{fig:GG2}, in which the integrals involve the propagator power configuration \(\vec{n} = (4, 1, 1, 1, 1, 1)\). In this case, we adopt a hybrid approach: I. use the Feynman parameterization method to compute the imaginary part of the first three loops ($\text{Im} B_{(4,1,1,1)} (s)$); II. substitute $\text{Im} B_{(4,1,1,1)} (s)$ into DR and perform the DR iteration method for the remaining two loops. Such a methodology can effectively circumvent the need for GDR. Since the results are rather complicated and lengthy, we provide an additional supplementary file ``CorrelationFunctions.pdf'' to show the expressions of correlation functions for both systems~\cite{supplementaryfile}.

\textit{\textbf{Numerical Analyses.}} \textemdash \,
In this section, we perform the QCD sum rule analyses for triple-$\eta_c$ and triple-$J/\psi$ systems by using the following values of various QCD parameters\cite{ParticleDataGroup:2024cfk,Ovchinnikov:1988gk,Narison:2018dcr}
\begin{equation}\label{parameters}
 \begin{aligned}
                    m_c&=1.27\pm 0.02\,\text{GeV}, \\
 \langle g_s^2G^2\rangle&=\langle0.48\pm0.14\rangle\,\text{GeV}^4.
 \end{aligned}
\end{equation}
To ensure the OPE convergence, we require the contribution of perturbative term to be three times that of the $\langle G^2\rangle$ term, thereby establishing a lower limit for the Borel parameter $M^2_B$
\begin{equation}
    \frac{\int_{36m_Q^2}^\infty e^{-s/M^2_B} \rho^{\text{Pert}}(s)\text{d}s}{\int_{36m_Q^2}^\infty e^{-s/M^2_B} \rho^{\langle G^2\rangle}(s)\text{d}s}\ge 3.
\end{equation}
\begin{figure}[htbp]
  \centering
  \subfigure[$3\eta_c$ system]{
    \includegraphics[width=0.43\textwidth]{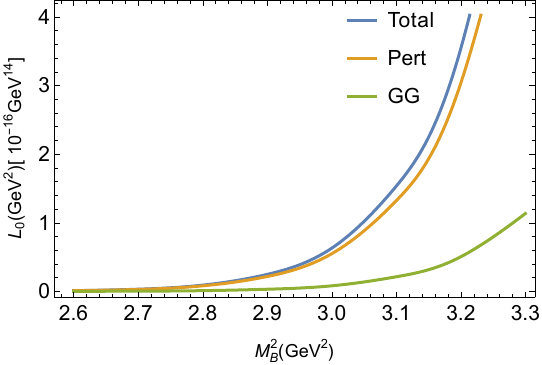}
    \label{fig:ope3etac}
  }
   \subfigure[$3J/\psi$ system]{
    \includegraphics[width=0.43\textwidth]{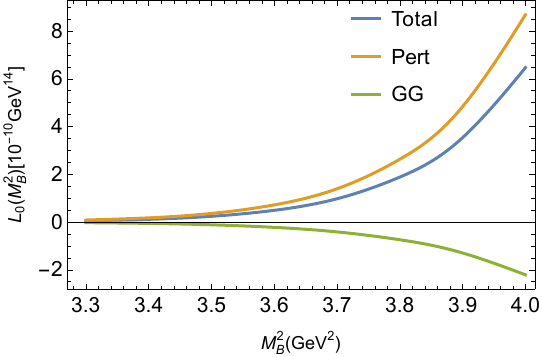}
    \label{fig:ope3Jpsi}
  }
  \caption{OPE convergences for the $3\eta_c$ (a) and $3J/\psi$ (b) systems.}
  \label{fig:ope}
\end{figure}
   \begin{figure}[htbp]
  \centering
  \subfigure[$3\eta_c$ system]{
    \includegraphics[width=0.43\textwidth]{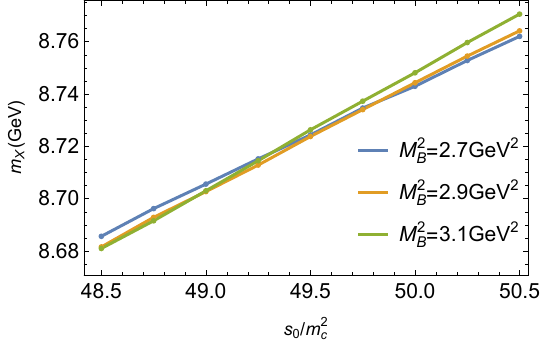}
    \label{fig:s01}
  }
   \subfigure[$3J/\psi$ system]{
    \includegraphics[width=0.43\textwidth]{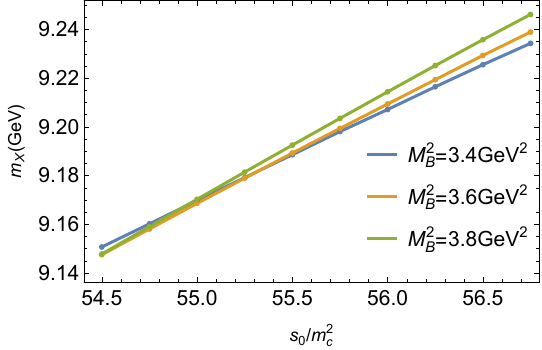}
    \label{fig:s01}
  }
  \caption{Variations of hadron mass $m_X$ with $s_0/m^2_c$ for the $3\eta_c$ (a) and $3J/\psi$ (b) systems.}
  \label{fig:s0}
\end{figure}
For both triple-$\eta_c$ and triple-$J/\psi$ systems, we show the contributions of the perturbative terms and gluon condensates in Fig.~\ref{fig:ope}, in which the perturbative terms are much bigger than the gluon condensates. In multiquark systems, the dimension-six tri-gluon condensates typically contribute significantly less than the dimension-four gluon condensates. The tri-gluon condensate contributions have negligible effects and are therefore ignored in our calculations.

\begin{figure}[htbp]
  \centering
   \subfigure[$3\eta_c$ system]{
    \includegraphics[width=0.45\textwidth]{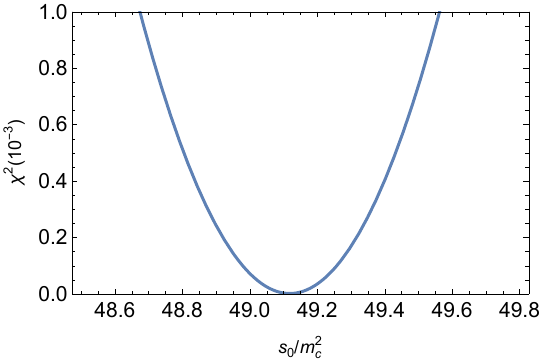}
    \label{fig:sub1}
  }
   \hfill  
     \subfigure[$3J/\psi$ system]{
    \includegraphics[width=0.45\textwidth]{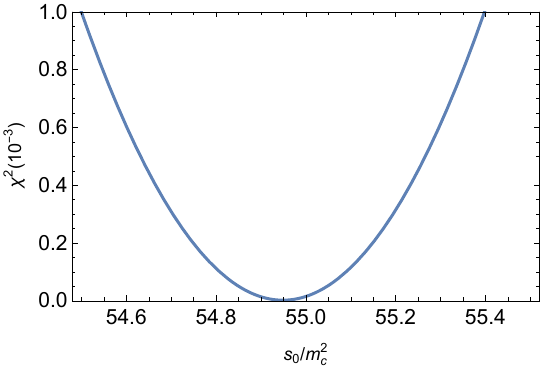}
    \label{fig:sub2}
  }
  \caption{The $\chi^2({s_0}/{m^2_c})$ for the $3\eta_c$ and $3J/\psi$ systems.}
  \label{fig:chi}
\end{figure}

After determining $M^2_{B\text{min}}$, we plot the $m_X - s_0/m^2_c$ curves for different values of $M^2_B$ in Fig.~\ref{fig:s0} to determine the optimal value of the continuum threshold $s_0$, around which the dependence of $m_X$ on $M^2_B$ is minimized. To achieve this, we define the function $\chi^2$ as
\begin{equation}
\chi^2(s_0/m^2_c)=\sum_{i=1}^N\left[\frac{m_X(s_0/m^2_c,M_{B,i}^2)}{\overline m_X(s_0/m^2_c)}-1\right]^2 , \label{eq:chi}
\end{equation}
where $\overline{m}_X(s_0/m_c^2)$ is the average of the data points
\begin{equation}
    \overline m_X(s_0/m^2_c)=\sum_{i=1}^N\frac{ m_X(s_0/m^2_c,M_{B,i}^2)}{N}.
\end{equation}
We show $\chi^2(s_0/m_c^2)$ in Fig.~\ref{fig:chi} to yield an optimal $s_0$ corresponding to the minimal point of the curves.

Subsequently, the upper limit for $M^2_B$ is established by ensuring that the pole contribution (PC) exceeds 40\%
\begin{equation}
    \text{PC}(s_0,M_B^2)=\frac{\mathcal{L}_0(s_0,M_B^2)}{\mathcal{L}_0(\infty,M_B^2)}\ge 40\%.
\end{equation}

\begin{figure}[htbp]
  \centering
   \subfigure[$3\eta_c$ system]{
    \includegraphics[width=0.45\textwidth]{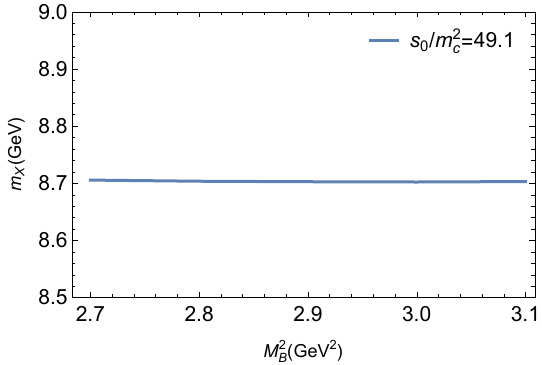}
    \label{fig:mass1}
  }
   \hfill  
     \subfigure[$3J/\psi$ system]{
    \includegraphics[width=0.45\textwidth]{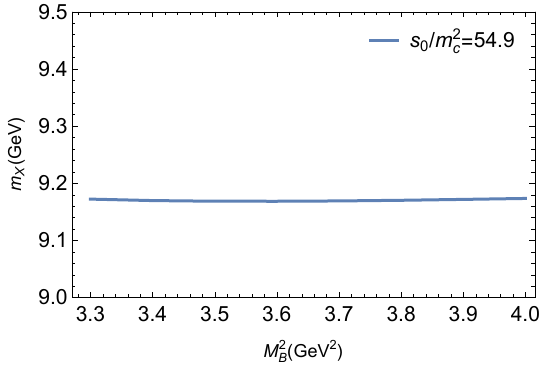}
    \label{fig:mass2}
  }
  \caption{Hadron mass $m_X$ for the $3\eta_c$ and $3J/\psi$ systems.}
  \label{fig:mass}
\end{figure}

With the chosen Borel window and $s_0$, we can establish very stable mass sum rules for the triple-$\eta_c$ and triple-$J/\psi$ systems. We show the Borel curves in Fig.~\ref{fig:mass} and determine the hadron masses for these two states. The numerical results are collected in Table~\ref{table:result}, in which the errors come from the charm quark mass $m_c$ and gluon condensate $\langle g^2_s G^2\rangle$ in Eq.~\eqref{parameters}. As shown in Figs.~\ref{fig:chi}-\ref{fig:mass}, the uncertainties from the parameters $s_0$ and $M_B$ are small enough to be neglected, as guaranteed by the $\chi^2$ approach and the stability of Borel curves. Meanwhile, we don't consider the uncertainties from the truncation of OPE series therefore the errors in Table~\ref{table:result} are actually underestimated.

\begin{table}[htbp] 
    \centering
    \renewcommand\arraystretch{1.4}
    \setlength{\tabcolsep}{1.8mm}
    {
        \begin{tabular}{cccccc}
            \hline\hline
        System & $J^{PC}$ & Mass[$\text{GeV}$] & $s_0/m^2_c$  & $M^2_B$[$\text{GeV}^2$] & $\text{PC}[\%]$\\
    $3\eta_c$ & $0^{-+}$ & $8.71 \pm 0.09$ & 49.1 & 2.7-3.1 & 47\\
    $3J/\psi$ & $1^{--}$ & $9.17 \pm 0.20$ & 54.9 & 3.4-3.8 & 43\\
            \hline\hline
        \end{tabular}
    }
    \caption{Numerical results for the $3\eta_c$ and $3J/\psi$ states.}\label{table:result}
\end{table}

The extracted masses for the triple-$\eta_c$ and triple-$J/\psi$ states are found to be below the corresponding three-meson mass thresholds $T_{3\eta_c}\approx 8.95$GeV and $T_{3J/\psi}\approx 9.29$GeV~\cite{ParticleDataGroup:2024cfk}, respectively. The binding energies are given as $E_B=0.24$GeV for the triple-$\eta_c$ and $E_B=0.12$GeV for the triple-$J/\psi$ systems.
This result supports the investigation in the Gaussian expansion method that there is a bound state of triple-$J/\psi$~\cite{Pan:2024ple}.

\textit{\textbf{Summary.}} \textemdash \,
In this work, we investigate the fully heavy triple-$\eta_c$ state with $J^{PC}=0^{-+}$ and triple-$J/\psi$ state with $J^{PC}=1^{--}$ and calculate their masses using the method of QCD sum rules. We calculate the two-point correlation functions and spectral functions by including the perturbative term and nonperturbative dimension-four gluon condensate. To improve computational efficiency, we employ an IDR approach to deal with the five-loop banana integrals appearing in the OPE calculations, and adopt two key techniques to avoid potential small-circle subtraction in the GDR representation. 

Establishing the stable mass sum rules, we extract hadron masses $M_{3\eta_c}=8.71\pm0.09\text{GeV}$ and $M_{3J/\psi}=9.17\pm0.20\text{GeV}$, which are below the corresponding three-meson mass thresholds for the triple-$\eta_c$ and triple-$J/\psi$ systems, respectively. Our results support the existence of the triple-$\eta_c$ and triple-$J/\psi$ bound states, in agreement with the prediction in Ref.~\cite{Pan:2024ple}. Moreover, a deeper bound state is suggested in the triple-$\eta_c$ system than that in the triple-$J/\psi$ system. With the CMS observation of three $J/\psi$ mesons production~\cite{CMS:2021qsn}, we suggest the possible measurement of such three-body states in future proton-proton collision processes. 

\textit{\textbf{Acknowledgement.}} \textemdash \,

Guo-Feng Xu thanks Zhi-Zhong Chen and Si-Yi Chen for valuable discussions. This work is supported by the National Natural Science Foundation of China under Grant No. 12175318 and the Natural Science Foundation of Guangdong Province of China under Grant No. 2023A1515011704.

\end{document}